\documentclass{article}
\usepackage{arxiv}

\usepackage[utf8]{inputenc} 
\usepackage[T1]{fontenc}    

\usepackage{url}            
\usepackage{graphicx}
\usepackage[hidelinks]{hyperref}       

\usepackage[backend=bibtex]{biblatex}
\bibliography{bibliography}

\usepackage{subcaption}
\usepackage[group-separator={,}]{siunitx}
\DeclareGraphicsExtensions{.pdf}
\usepackage{glossaries}
\makeglossaries
\newacronym{api}{API}{Application Programming Interface}
\newacronym{ceu}{CEU}{Central European University}
\newacronym{cdr}{CDR}{Call Detail Record}
\newacronym{csv}{CSV}{Comma Separated Value}
\newacronym{etl}{ETL}{extract, transform, load}
\newacronym{gis}{GIS}{Geographic Information System}
\newacronym{gps}{GPS}{Global Positioning System}
\newacronym{huf}{HUF}{Hungarian forint} 
\newacronym{imei}{IMEI}{International Mobile Equipment Identity}
\newacronym{ksh}{KSH}{Központi Statisztikai Hivatal, Hungarian Central Statistical Office}
\newacronym{mti}{MTI}{Magyar Távirati Iroda, Hungarian news agency}
\newacronym{osm}{OSM}{OpenStreetMap}
\newacronym{ses}{SES}{Social Economic Status}
\newacronym{sim}{SIM}{Subscriber Identity Module}
\newacronym{tac}{TAC}{Type Allocation Code}
\newacronym{umts}{UMTS}{Universal Mobile Telecommunications System}
\newacronym{wgs84}{WGS 84}{World Geodetic System, also known as EPSG:4326}
\newacronym{wkt}{WKT}{Well-known text}

\usepackage[titletoc,title]{appendix}

\begin{document}
\title{Analyzing the Behavior of Soccer Fans from a Mobile Phone Network Perspective: Euro 2016, a Case Study}

\author{Gerg\H{o} Pint\'{e}r \\
John von Neumann Faculty of Informatics, \\
\'{O}buda University, \\
B\'{e}csi \'{u}t 96/B, 1034 Budapest, Hungary \\
\texttt{pinter.gergo@uni-obuda.hu} \And
Imre Felde \\
John von Neumann Faculty of Informatics,\\
\'{O}buda University,\\
B\'{e}csi \'{u}t 96/B, 1034 Budapest, Hungary \\
\texttt{felde.imre@uni-obuda.hu}
}

\keywords{mobile phone data; call detail records; data analysis; human mobility; large social event; social sensing}

\maketitle{}

\begin{abstract}
In this study, we analyzed one month of Call Detail Records, covering Budapest in June, 2016. During this observation period, the 2016 UEFA European Football Championship took place, where the Hungarian national football team did well, beyond expectations. This case study focuses on the fans' behavior in Budapest, during and after the Hungarian matches, from the perspective of the mobile phone network data.
Tense rooting, joy of scoring a goal, self-forgetting celebration because of winning the group.
All of these have an effect on the data, that can be simply revealed.
We used descriptive statistical methods to analyze the available data set, and managed to demonstrate the potential of this kind of data at social sensing.
We found that the mobile phone network activity reflects the football fans' behavior, even if the match is played in another country.
\end{abstract}

\section{Introduction}
\label{sec:introduction}

Football is a popular sport, European or World Championships, especially the finals, are among the most watched sporting events. The Euro 2016 Final was watched by more than 20 million people in France \cite{variety2016soccer}, or the
Germany vs. France semifinal was watch by almost 30 million people in Germany \cite{variety2016soccer}. But what about Hungary?

According to the MTVA (Media Services and Support Trust Fund), that operates the television channel M4~Sport, the first Hungarian match was watched by about 1.734 million people, the second by about 1.976 million and the third group match by about 2.318 million people\footnote{According to the Hungarian Central Statistical Office (\acrshort{ksh}), the population of Hungary was about 9.83 million in 2016 \cite{ksh22.1.1.1}.}. With these ratings, the M4~Sport, became the most watched television channel in Hungary, on those days \cite{hiradohu2016csoportgyoztes}.

The whole participation of the Hungarian national football team was beyond expectations and raised interest, even among those, who generally, do not follow football matches.
In the beginning, it might have been because Hungary returned to the European Championship after 44 years. Later, a good performance by the national football team, increased the interest.
But, is it possible to measure/correlate this interest, with a mobile phone network?

In this study, we analyzed the mobile phone network activity before, during and after the matches of the Hungarian national football team, but not directly at the location of the matches. The Call Detail Records (\acrshort{cdr}) analyzed in this paper, covers Budapest, the capital of Hungary. We present another example of social sensing using \acrshort{cdr}s, in an indirect and a direct way. Indirectly, as the mobile phone activity of the sport fans, residing in Budapest, are studied during a matches played in France. Directly, as the spontaneous festival on the streets of Budapest, after the third match, is presented from a data perspective.

The rest of this paper is organized as follows. After a brief literature review in Section~\ref{sec:literature_review}, the utilized data is described in Section~\ref{sec:data}, then, in Section~\ref{sec:results} the results of this case study are introduced. Finally, in Section~\ref{sec:discussion}, the findings of the paper are summarized.

\section{Literature Review}
\label{sec:literature_review}

Mobile phones can work as sensors, that detect the whereabouts and movement of their carrier. In this day and age, practically everyone has a mobile phone, that makes it possible to use large scale analyses. With enough data, the general mobility customs can also be studied. The home and work locations can be determined \cite{pappalardo2021evaluation}, and based on those locations, the commuting trends can be identified and validated with census data \cite{pinter2021evaluating}.

Mobility indicators, such as Radius of Gyration or Entropy, are often calculated \cite{pappalardo2015returners,xu2018human} to describe and classify the subscribers' mobility customs. Furthermore, using mobility to infer about socioeconomic status is current direction of mobility analysis \cite{xu2018human,pinter2021evaluating}

\acrshort{cdr} analysis is often used \cite{traag2011social,xavier2012analyzing,mamei2016estimating,marques2018understanding,pinter2019activity,rotman2020using,hiir2020impact} for large social event detection.
When thousands of people are on the same place at the same time, they generate a significant `anomaly' in the data, whereas small groups usually do not stand out from the `noise'. This is especially true when the passive, transparent communication between the mobile phone device and the cell are not included in the data, but only the active communication (voice calls, text messages and data transfer) are recorded.

In \cite{pinter2019activity} and \cite{rotman2020using}, mass protests are analyzed via mobile phone network data.
In \cite{traag2011social}, \cite{mamei2016estimating}, \cite{xavier2012analyzing} and \cite{hiir2020impact}, the authors examined the location of stadiums, where the a football matches took place. Traag et al. \cite{traag2011social} and Hiir et al. \cite{hiir2020impact} also found that the mobile phone activity of the attendees decreased significantly, also used z-score to express the activity deviation during the social event from the average \cite{traag2011social}. Xavier et al. compared the reported number of attendees with the detected ones.
These works also analyze other social events like concerts and festivals.
However, in this paper, the actual matches took place in another country (France), and the local fans are examined.

Mobile phone network data is also used to analyze the human mobility during COVID-19 pandemic and the effectiveness of the restrictions.
Willberg et al. identified a significant decrease of the population presence in the largest cities of Finland after the lockdown compared to a usual week \cite{willberg2021escaping}.
Bushman et al. analyzed the compliance to social distancing in the US using mobile phone data \cite{bushman2020effectiveness}. Gao et al. found negative correlation in stay-at-home distancing and COVID-19 increase rate \cite{gao2020association}.
Still, these analyses might not be common enough. Oliver et al. asked the question: `Why is the use of mobile phone data not widespread, or a standard, in tackling epidemics?' \cite{oliver2020mobile}. This, however, is not within the scope of this paper.

\section{Data}
\label{sec:data}

Vodafone Hungary, one of the three mobile phone operators providing services in Hungary, provided anonymized \acrshort{cdr} data for this study. The observation area was Budapest, capital of Hungary and its agglomeration, and the observation period is one month (June, 2016). In 2016 Q2, the nationwide market share of Vodafone Hungary was 25.3\% \cite{nmhh_mobile_market_report}. This data set contains \num{2291246932} records from \num{2063005} unique \acrshort{sim} cards, and does not specify the type of the activity (voice call, text message, data transfer).

\begin{figure}[ht]
    \centering
    \includegraphics[width=\linewidth]{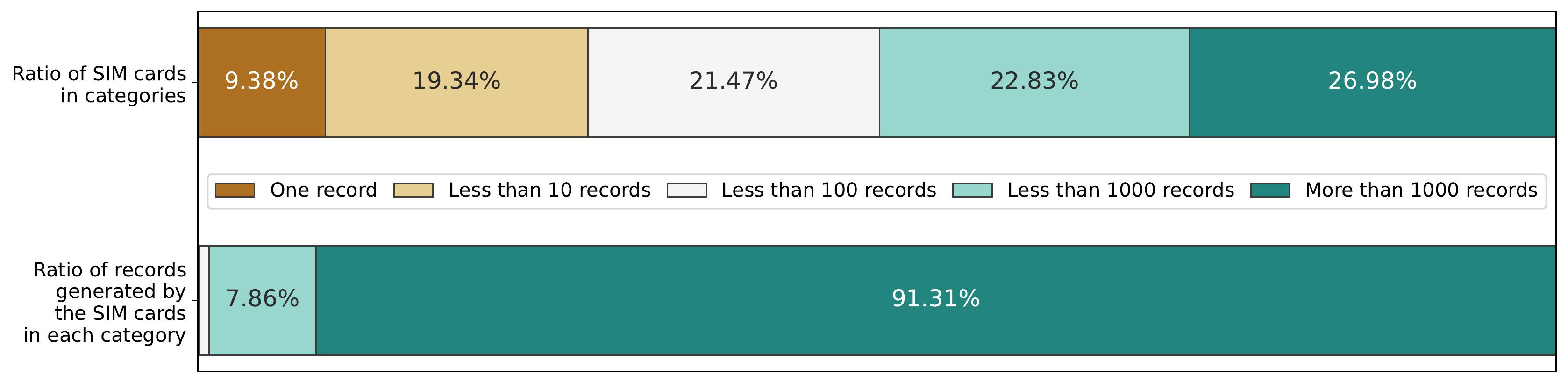}
    \caption{\acrshort{sim} cards categorized by the number of activity records.
    The \acrshort{sim} cards with more than 1000 activity records (26.98\% of the \acrshort{sim} cards) provide the majority (91.31\%) of the activity.}
    \label{fig:vod201606_sim_activity}
\end{figure}

Figure~\ref{fig:vod201606_sim_activity}, shows the activity distribution between the activity categories of the \acrshort{sim} cards. The dominance of the last category, the \acrshort{sim} cards with more than 1000 activity records, is even more significant. This almost 27\% of the \acrshort{sim} cards produce the more the 91\% of the activity.

\begin{figure}[ht]
    \centering
    \includegraphics[width=\linewidth]{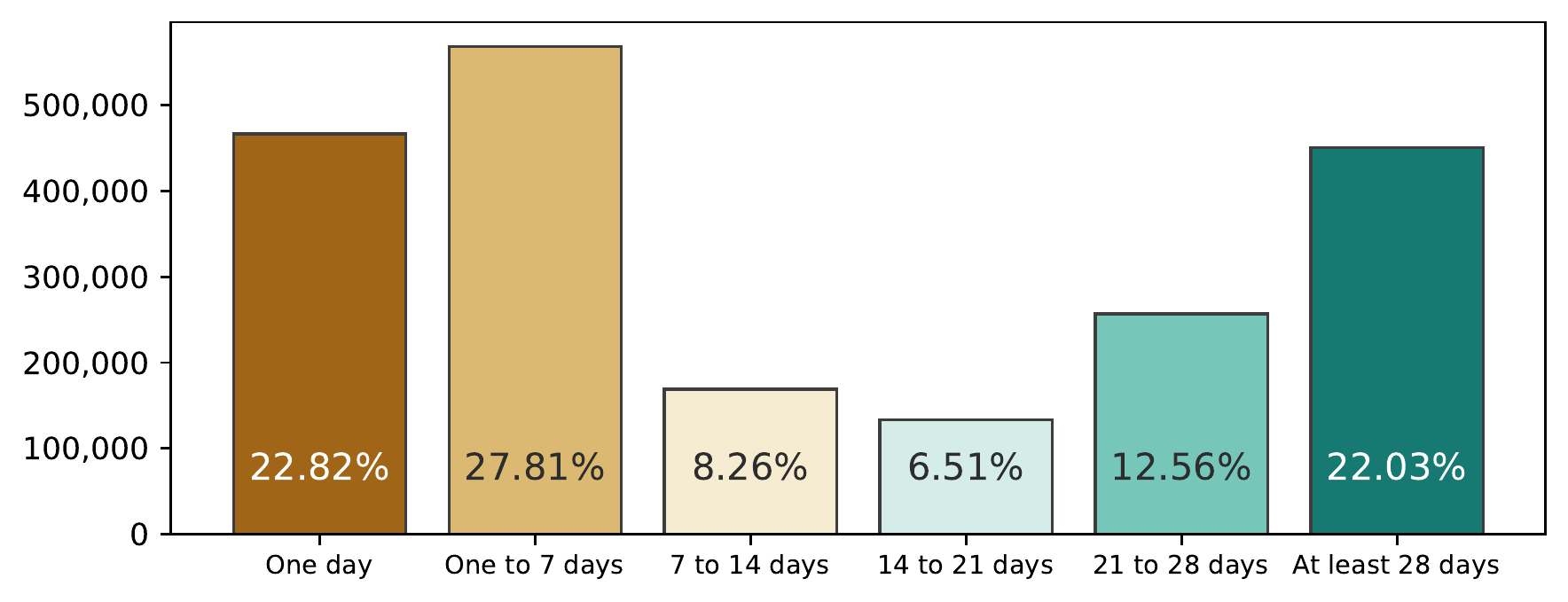}
    \caption{\acrshort{sim} card distribution by the number of active days.}
    \label{fig:vod201606_activity_by_days}
\end{figure}

Figure~\ref{fig:vod201606_activity_by_days}, shows the \acrshort{sim} card distribution by the number of active days. Only the 34.59\% of the \acrshort{sim} cards have activity on at least 21 different days.
There are \num{241824} \acrshort{sim} cards (11.72\%), that has appearance at least two days, but the difference between the first and the last activity is not more the seven days. This may indicate the presence of tourists, that is usual in this part of the year.

The received data is in a `wide' format, where all of fields are present for every record, that contains a
\acrshort{sim} ID, a timestamp, cell ID, the base station (site) coordinates in \acrshort{wgs84} projection, the subscriber (age, sex) and subscription details (consumer/business and prepaid/postpaid). While the subscription details are available for every \acrshort{sim} cards, the subscriber information is missing in slightly more than 40\% of the cases, presumably because of the subscribers' preferences of personal data usability.

\begin{figure}[ht]
    \centering
    \includegraphics[width=\linewidth]{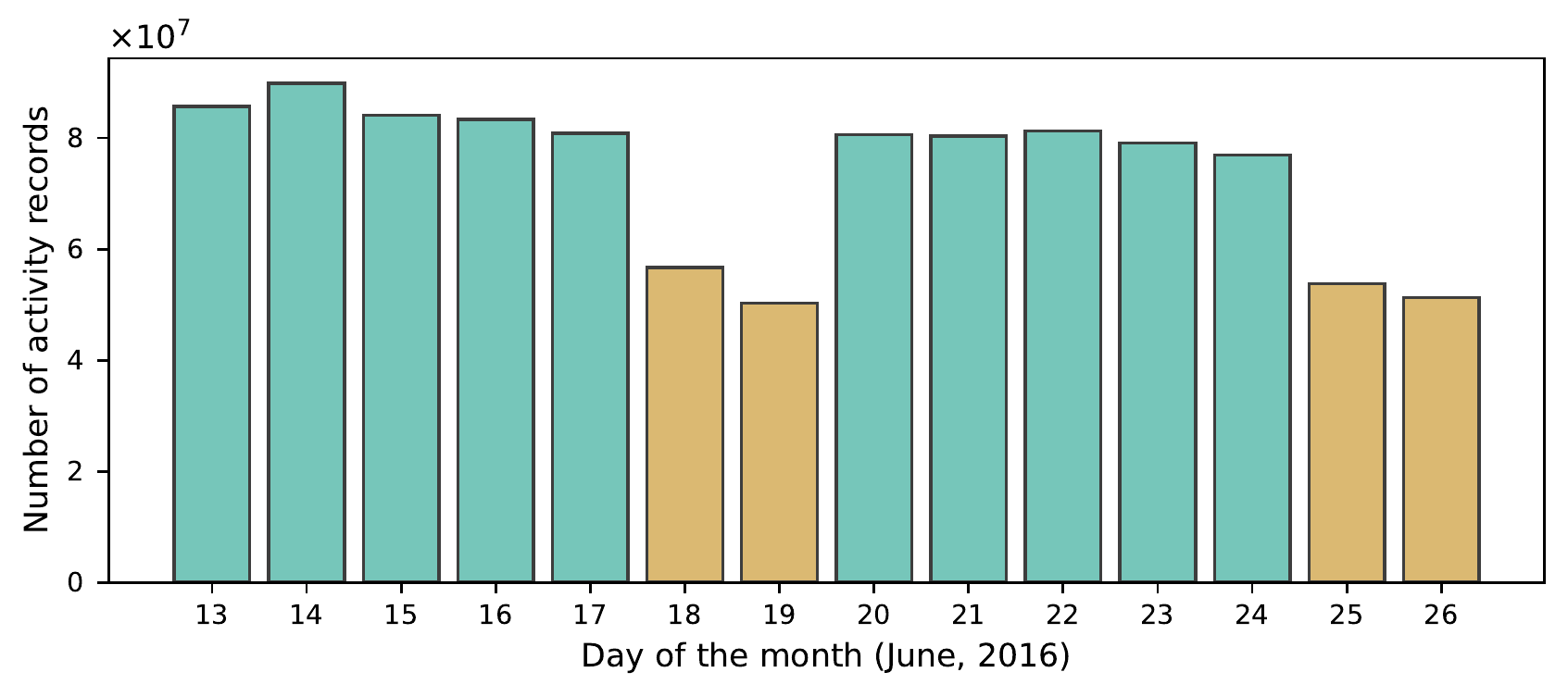}
    \caption{Number of daily activity records, during two weeks of June, 2016. The matches of the Hungarian national football team took place from June 14 to June 26.}
    \label{fig:vod201606_daily_activity}
\end{figure}

Figure~\ref{fig:vod201606_daily_activity}, shows number of daily activity records during the second half of the month. Weekends (brown bars) show significantly fewer activity, hence the activity during the matches compared to the weekday or weekend activity average, according to the day of the match.

Although the data contains cell IDs, only the base station locations are known, where the  cell antennas are located.
As a base station usually serve multiple cells, these cells has been merged by the serving base stations. After the merge, 665 locations (sites) remained with known geographic locations. To estimate the covered area of these sites, the Voronoi tessellation, has been performed on the locations, that is common practice \cite{pappalardo2016analytical,csaji2013exploring,vanhoof2018comparing,candia2008uncovering,novovic2020uncovering,trasarti2015discovering} for \acrshort{cdr} processing.

\section{Results}
\label{sec:results}

This section , Budapest downtown was analyzed spatially, and the \acrshort{cdr}s were filtered temporally, to select the match durations including two hours before and after the matches. This section discusses the results in the order of the Hungarian Euro 2016 matches.

\subsection{Austria vs. Hungary}

\begin{figure}[ht]
    \centering
    \includegraphics[width=\linewidth]{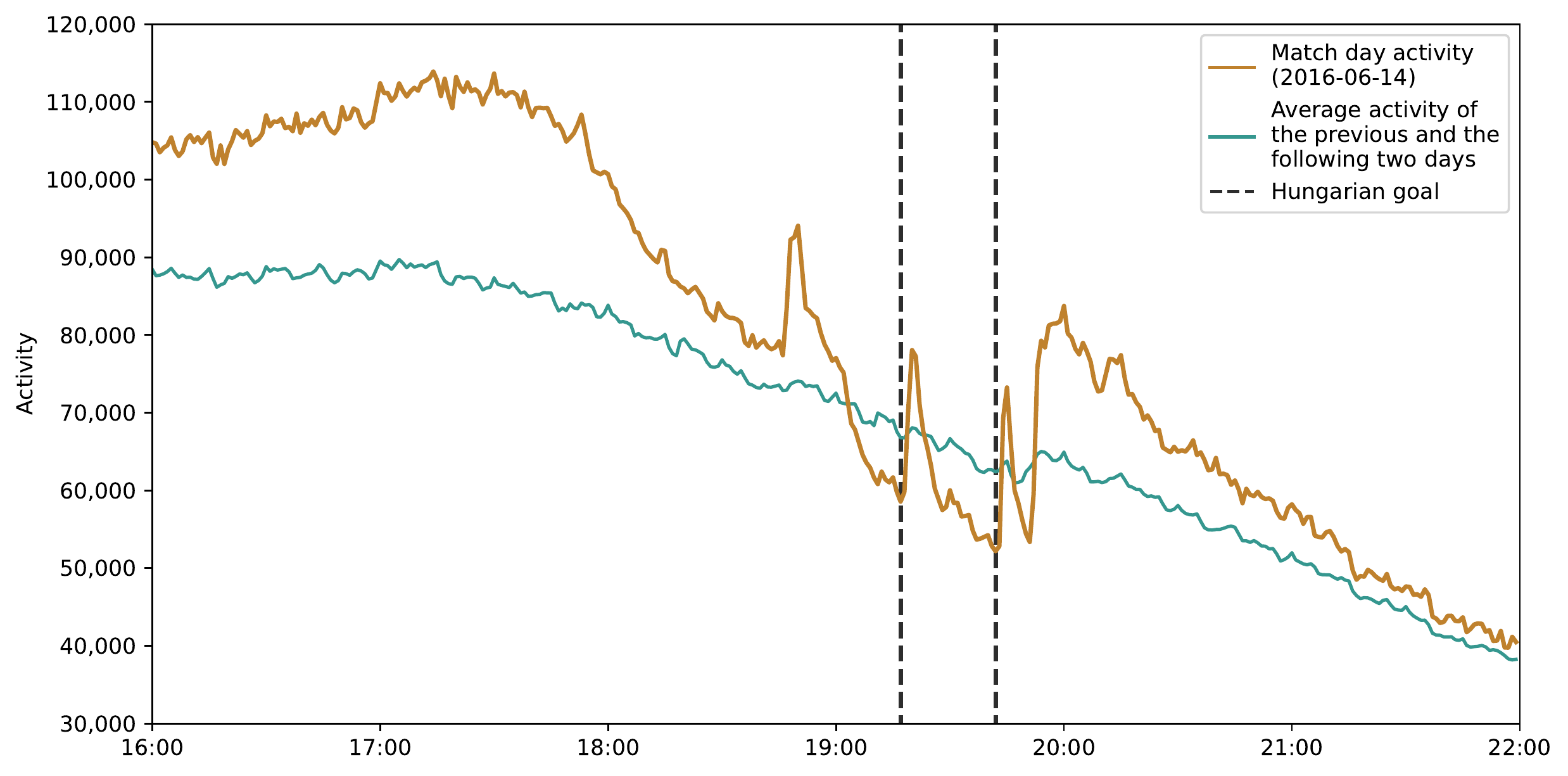}
    \caption{Mobile phone activity during and after the Austria--Hungary Euro 2016 match, in comparison with the average activity of the weekdays.}
    \label{fig:aut_hun_timeseries}
\end{figure}

The first match was against Austria (Figure~\ref{fig:aut_hun_timeseries}) on Tuesday, June 14, 2016. Before the match, the activity level is significantly higher than the average of the weekdays, then decreases until the half-time. During the second half, the activity level went below the average, as if more and more people started to watch the match and cease their other activities. Right after the Hungarian goals, there are two peaks in the activity.

Unfortunately, the data source cannot distinguish the mobile phone activities by type, so it cannot be known what kind of activities caused the peaks. It is supposed that mostly data transfer, maybe text messages, instead of phone calls. It simply does not seem to be lifelike to call someone during the match just because of a goal, but sending a line via one of the popular instant messaging services is feasible.

\subsection{Iceland vs. Hungary}

\begin{figure}[ht]
    \centering
    \includegraphics[width=\linewidth]{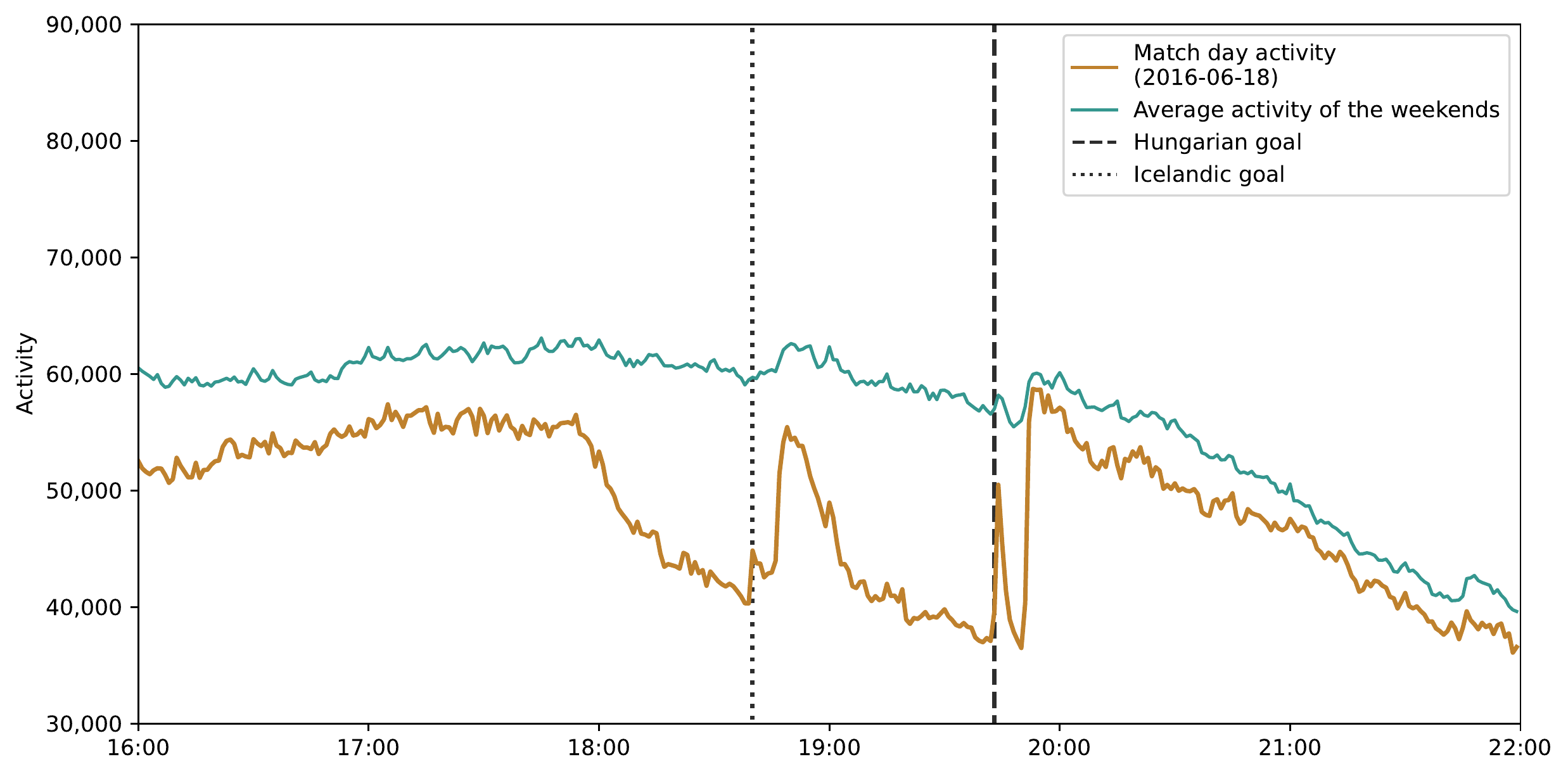}
    \caption{Mobile phone activity during and after the Iceland--Hungary Euro 2016 match, in comparison with the average activity of the weekends.}
    \label{fig:isl_hun_timeseries}
\end{figure}

The second match was against Iceland on Saturday, June 18, 2016. Figure~\ref{fig:isl_hun_timeseries}, shows the mobile phone activity levels before, during and after the match. As the weekend activity is generally lower (see Figure~\ref{fig:vod201606_daily_activity}), the average of the weekdays are used as a reference. The match began at 18:00, and from that point, the activity level is significantly below the average, except the half-time break and, again, the peak after the Hungarian goal.
Interestingly, the Icelandic goal does not result such a significant peak, only a very moderate one can be seen in the time series.

\subsection{Hungary vs. Portugal}

\begin{figure}[ht]
    \centering
    \includegraphics[width=\linewidth]{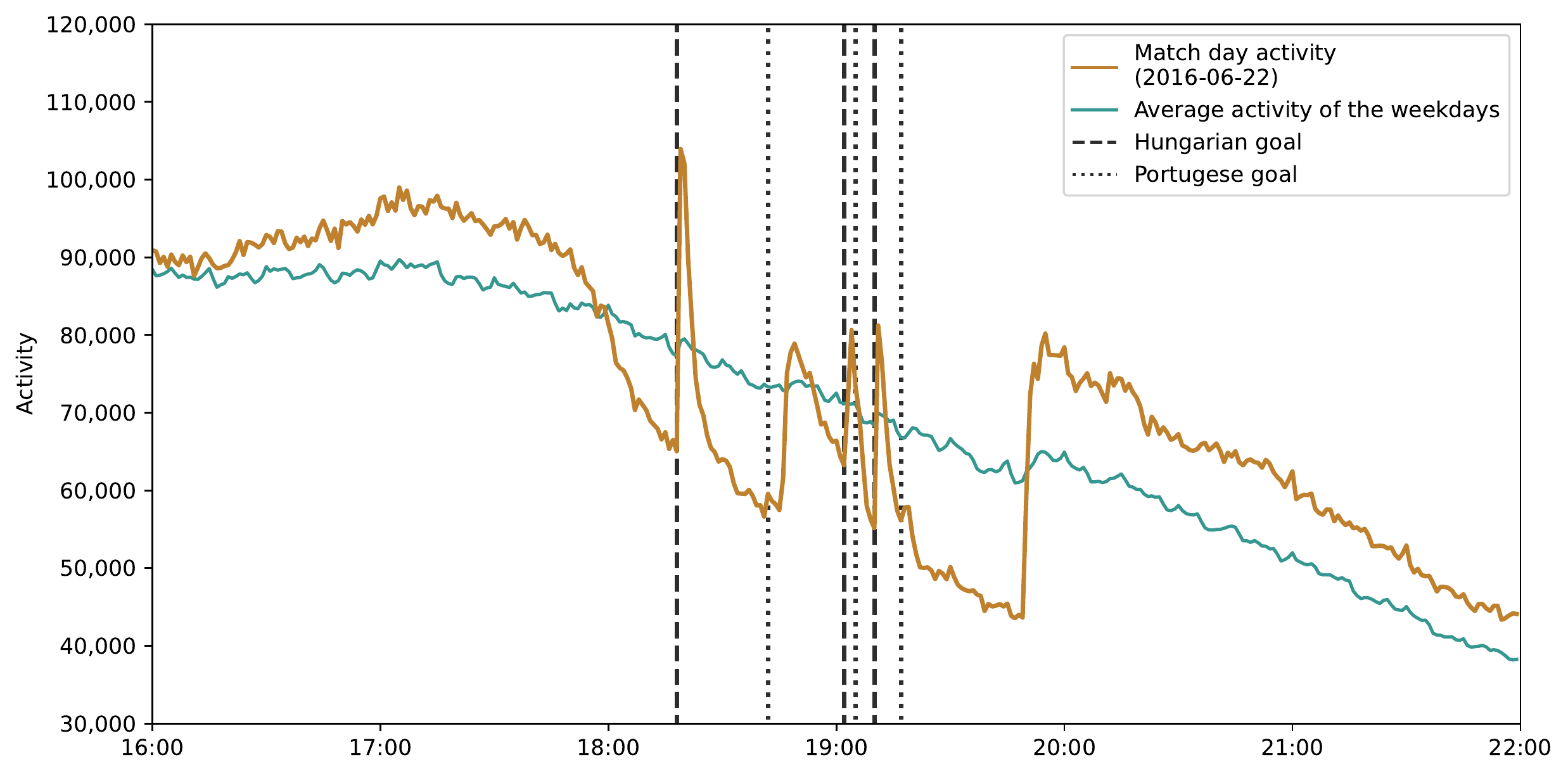}
    \caption{Mobile phone activity during and after the Hungary--Portugal Euro 2016 match, in comparison with the average activity of the weekdays.}
    \label{fig:hun_prt_timeseries}
\end{figure}

On Wednesday, June 22, 2016, as the third match of the group state of the 2016 UEFA European Football Championship, Hungary played draw with Portugal.
Both teams scored three goals and with this result, Hungary won their group and qualified for the knockout phase. 
During the match, the mobile phone activity dropped below the average, as before, but the goals against Portugal resulted significant peaks, especially the first one (see Figure~\ref{fig:hun_prt_timeseries}). The Portuguese equalizer goal(s) did not cause significant mark in the activity. In the second half, the teams scored four goals in a relatively short time period, but only the Hungarian ones resulted peaks.

\begin{figure}[ht]
    \centering
    \includegraphics[width=\linewidth]{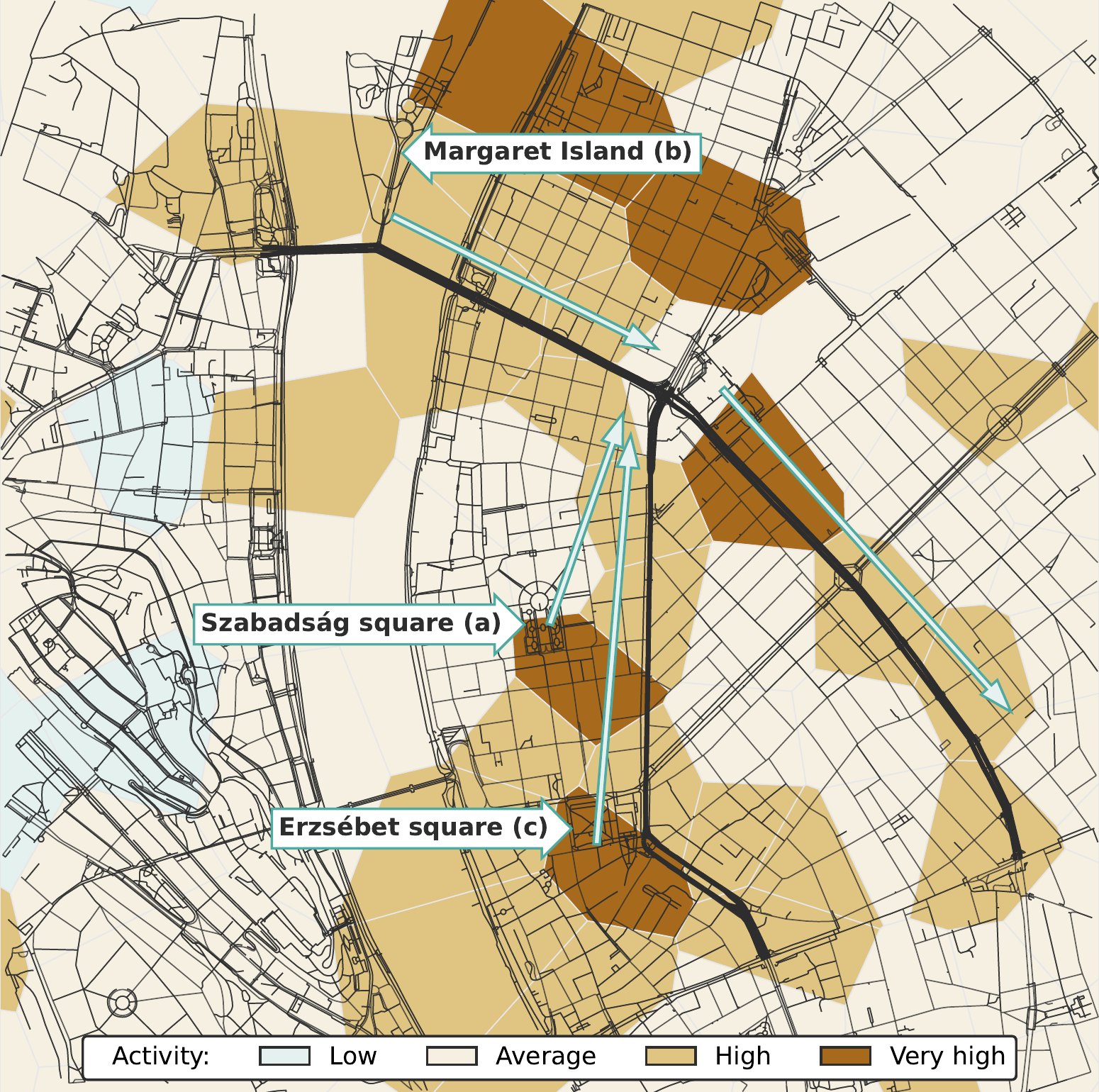}
    \caption{After the Hungary vs. Portugal football match, the fans, delirious with joy, filled the streets. The arrows show their route from the main fan zones to and along the Grand Boulevard. Voronoi polygons colored by the mobile phone network activity at the peak of the event, at 20:20.}
    \label{fig:post_match_festival}
\end{figure}

After the match, the activity level is over the average, that may represent the spontaneous festival in Budapest downtown. According to the \acrshort{mti} (Hungarian news agency), thousands of people celebrated on the streets, starting from the fan zones, mainly from the Erzsébet square (Figure~\ref{fig:post_match_festival} a), the Margaret Island (Figure~\ref{fig:post_match_festival} b) and Erzsébet square (Figure~\ref{fig:post_match_festival} c) towards Budapest Nyugati railway station. The Grand Boulevard was completely occupied and the public transportation was disrupted along the affected lines.

\begin{figure}[ht]
    \centering
    \includegraphics[width=\linewidth]{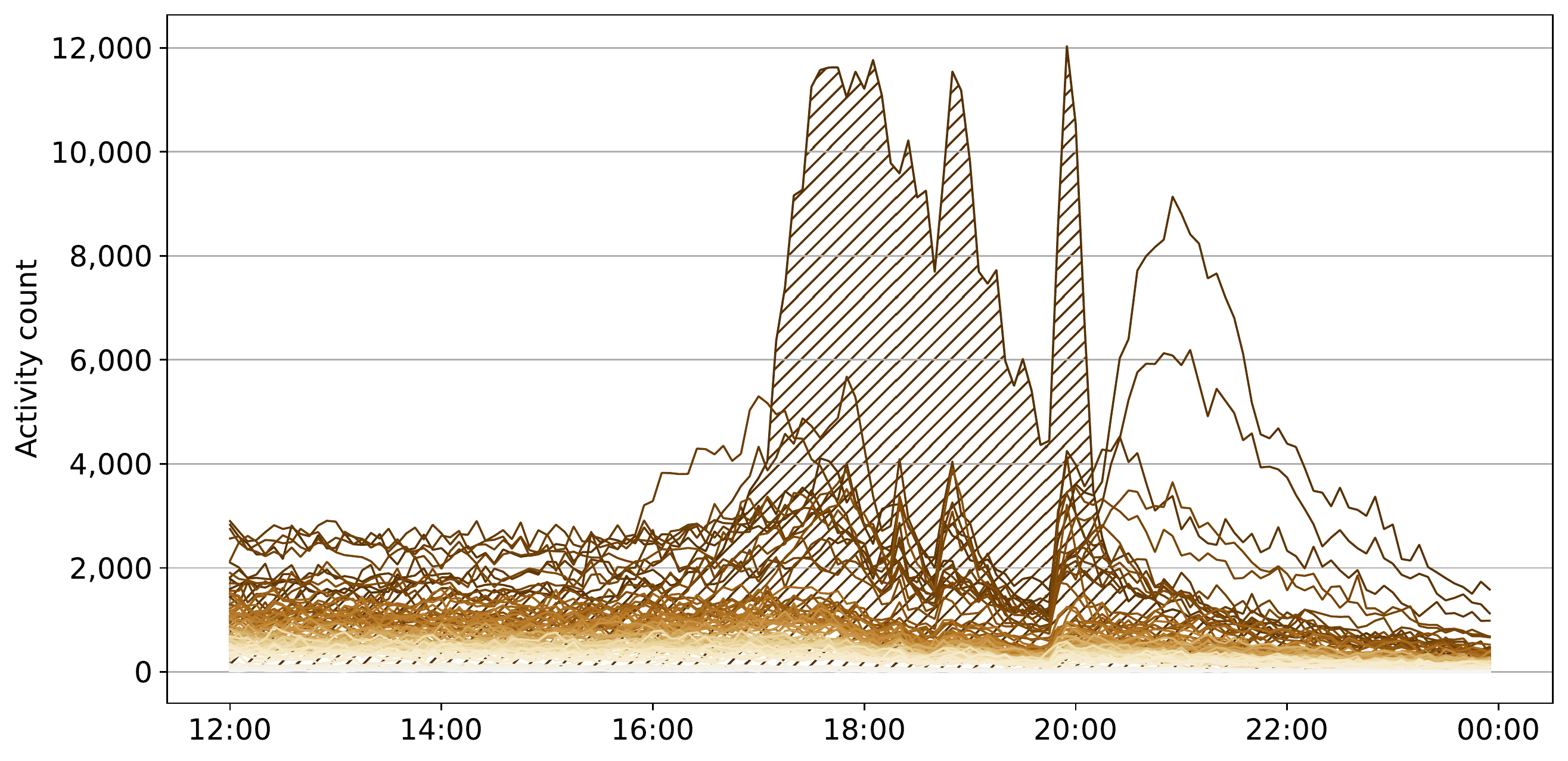}
    \caption{Site activities, in Budapest downtown, on the day of the Hungary vs. Portugal football match (June 22, 2016). The highlighted site covers mostly the Szabadság square, where one of the main fan zone was set up.}
    \label{fig:post_match_festival_timeseries}
\end{figure}

Figure~\ref{fig:post_match_festival_timeseries}, shows the activities of the sites (multiple cells aggregated by the base stations) in Budapest downtown.
The highlighted site covers mostly the Szabadság square (for the location, see Figure~\ref{fig:post_match_festival} a), where one of the main fan zone was set up, with big screen and so on. The activity curve actually follows the trends of the whole data set (see Figure~\ref{fig:hun_prt_timeseries}). There is high activity before the match, during half-time and, for a short period, after the match. During the match, the activity decreases with the except of four not so significant peaks around the goals.

In the highlighted site, in Figure~\ref{fig:post_match_festival_timeseries}, almost 10 thousand \acrshort{sim} cards had been detected between 17:00 and 20:00. 50.26\% of the subscribers were between 20 and 50 years old, 35.8\% of them, had no age data.

After the match, there is a significant increase in the activity in some other sites. These sites are (mostly) around the Grand Boulevard, where the fans marched, celebrating the advancement of the national football team, to the knockout phase.

Figure~\ref{fig:post_match_festival}, shows spatial distribution of this social event, using Voronoi polygons generated around the base stations locations.
The polygons are colored by the mobile phone network activity increase, compared to average of the weekday activity, at 20:20. For the comparison, the standard score
\footnote{The standard score (or z-score) is defined as ${z = \frac{x-\mu}{\sigma}}$, where $\mu$ is the mean and $\sigma$ is the standard deviation.}
was determined for every base station with a 5-minute temporal aggregation. The darker colors indicate the higher activity surplus in an area.
For the details of determining the thresholds, see Appendix~\ref{app:zscore_thresholds}.
The figure also denotes the three main fan zones in the area, route of the fans by arrows, and the affected streets are emphasized.

\subsection{Hungary vs. Belgium}

\begin{figure}[ht]
    \centering
    \includegraphics[width=\linewidth]{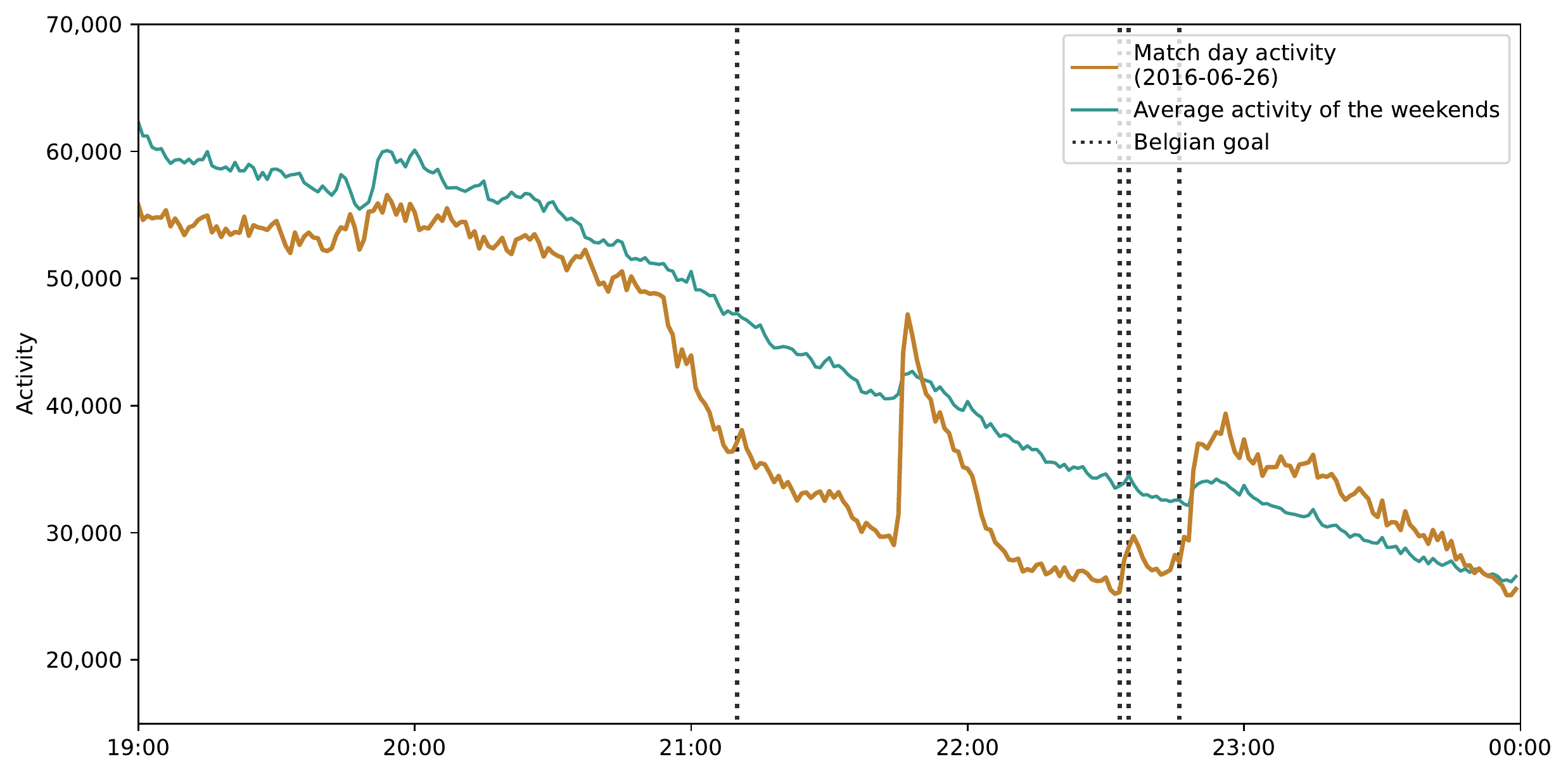}
    \caption{Mobile phone activity during and after the Hungary--Belgium Euro 2016 match, in comparison with the average activity of the weekends.}
    \label{fig:hun_bel_timeseries}
\end{figure}

On Sunday, June 26, 2016, Hungary played the fourth and last Euro 2016 match against Belgium. Figure~\ref{fig:hun_bel_timeseries}, shows the mobile phone network activity before, during and after the match.
During the match, the activity level was below the weekend average.
Although, after the match, the activity is slightly higher than the average, but the match ended late (on Sunday), when the activity average is very low. This activity surplus may only indicate that the fans were simply leaving the fan zones and going home.

\section{Discussion}
\label{sec:discussion}

In this case study, we demonstrated that mobile phone network activity follows precisely the football fans' behavior, even if the matches are played in another country. So, in this analysis, people watched the matches on TV (at home) or big screens in the fan zones, but not in the stadium, where the matches were played.

The time series clearly show that the activity was below the average during the matches, indicating that many people did nothing other than rooting for their team. This coincide with other studies, where the activity of cells at the stadium was analyzed. However, this study does not focused on a small location like a stadium, but a large city, where people watched the matches on screens.
We managed to demonstrate that a remote football match can also have significant effect on the mobile phone network.
Moreover, the joy felt after the Hungarian goals, is clearly manifested in the data, as sudden activity peaks.

The spontaneous festival after the Hungary vs. Portugal match is, however, a direct application of social sensing and comparable to mass protests from a data perspective. During the event, the mobile phone network activity was significantly higher than the average in the affected areas.
We also presented an analysis of a fan zone, that is comparable to other studies, where smaller locations were analyzed. The mobile phone network activity at the fan zone, shows similar trend as whole data set. The activity decreases during the match, except for the half-time. There are also peaks right after the goals, although not so significant ones.

\vspace{6pt}

\section*{Author Contributions}
Conceptualization, G.P. and I.F.; methodology, G.P.; software, G.P.; validation, G.P.; formal analysis, G.P.; investigation, I.F.; resources, I.F.; data curation, G.P.; writing---original draft preparation, G.P.; writing---review and editing, G.P.; visualization, G.P.; supervision, I.F.; project administration, I.F.; funding acquisition, I.F. All authors have read and agreed to the published version of the manuscript.

\section*{Funding}
This research supported by the project 2019-1.3.1-KK-2019-00007 and by the Eötvös Loránd Research Network Secretariat under grant agreement no. ELKH KÖ-40/2020.

\section*{Acknowledgments}
The authors would like to thank Vodafone Hungary for providing the Call Detail Records for this study.

For plotting the map, OpenStreetMap data is used, that is copyrighted by the OpenStreetMap contributors and licensed under the Open Data Commons Open Database License (ODbL).

\section*{Conflicts of Interest}
The authors declare no conflict of interest. The funders had no role in the design of the study; in the collection, analyses, or interpretation of data; in the writing of the manuscript, or in the decision to publish the results.

\printglossary[title=Abbreviations, toctitle=Abbreviations, nogroupskip=true]

\begin{appendices}
\appendix
\section{}
\label{app:zscore_thresholds}

\begin{figure}[t!]
  \centering
  \includegraphics[width=\linewidth]{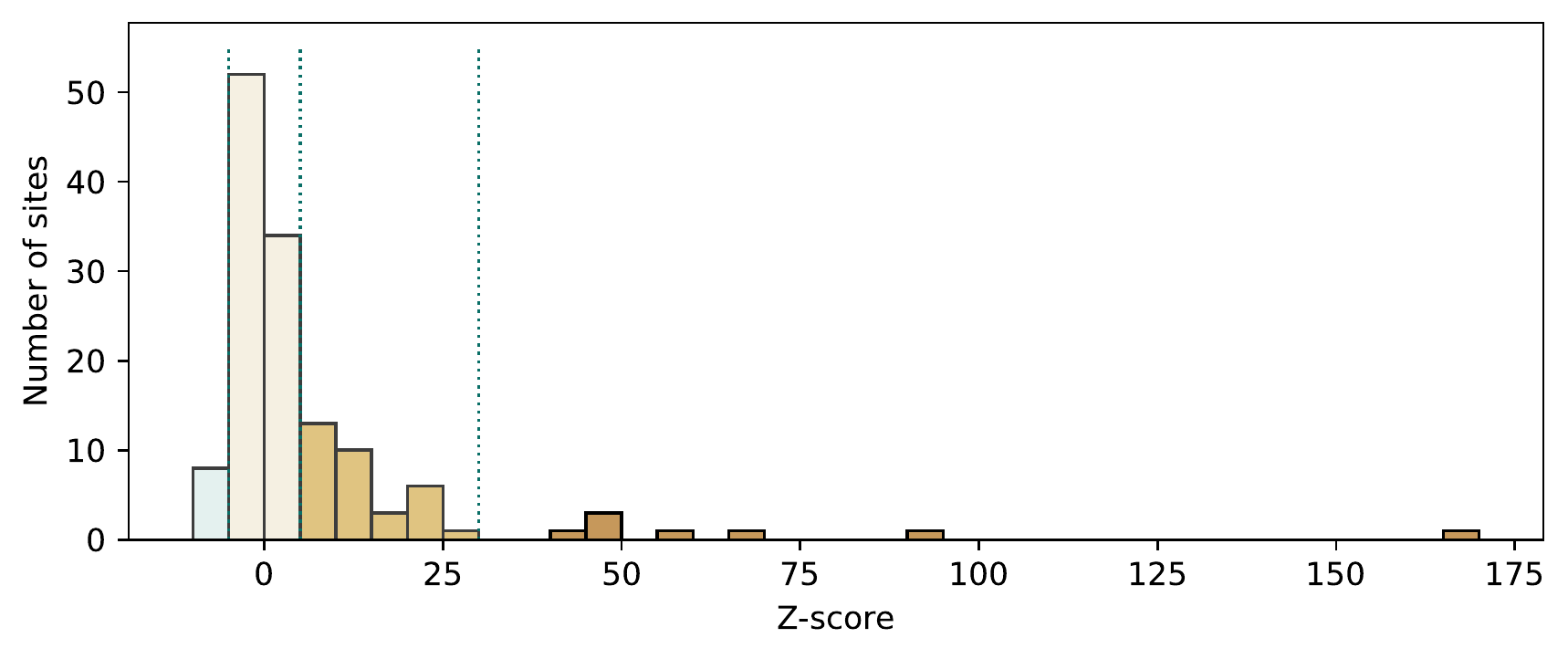}
  \caption{Z-score distribution of the downtown sites, with the activity level thresholds at $-5$, 5 and 30, using the same colors as in Figure~\ref{fig:post_match_festival}.}
  \label{fig:zscore_hist}
\end{figure}

To determine the activity levels for the map (Figure~\ref{fig:post_match_festival}), the match day activity, the average weekdays activity and the Z-scores are determined for the sites of the area of interest (downtown), in the selected time interval (20:15--20:20). Then, the histogram of the Z-score are generated for the selected sites (see Figure~\ref{fig:zscore_hist}) to determine the activity categories. Zero means that, the activity level is exactly the same as the average, but a wider interval (between $-5$ and $5$) is considered average to allow some variation.
Sites with Z-score between $5$ and $30$ are considered having high activity during the given time interval. There are sites with either low (below $-5$) or very high activity (over $30$).

\end{appendices}

\printbibliography

\end{document}